\documentclass[a4paper,11pt]{article}
\pdfoutput=1 

\usepackage{jinstpub} 

\title{\boldmath Components Qualification for a Possible use in the Mu2e Calorimeter Waveform Digitizer}

\author[a]{S.~Di Falco,}
\author[a, b]{S.~Donati,}
\author[a, c, 1]{L.~Morescalchi\note{Corresponding author.}}
\author[a]{E. Pedreschi,}
\author[a, b]{G.~Pezzullo,}
\author[a]{and F.~Spinella}

\affiliation[a]{INFN Pisa}
\affiliation[b]{University of Pisa}
\affiliation[c]{University of Siena}

\emailAdd{luca.morescalchi@pi.infn.it}

\abstract{The Mu2e experiment at Fermilab searches for the charged flavor violating  conversion of a muon into an electron in the Coulomb field of a nucleus. The detector consists of a straw tube tracker and a CSI crystal electromagnetic calorimeter, both housed in a superconducting solenoid. Both the front-end and the digital electronics, located inside the cryostat, will be operated in vacuum under a 1~T  magnetic field, having to sustain the high flux of neutrons and ionizing particles coming from the muons stopping target. These harsh experimental conditions make the design of the calorimeter waveform digitizer quite challenging. All the selected commercial devices must be tested individually and qualified for radiation hardness and operation in high magnetic field.
At the moment the expected particles flux and spectra at the digitizers location are not completely simulated and we are using initial rough estimates to select the components for the first prototype. We are gaining experience in the qualification procedures using the selected components but the choice will be frozen only when dose and neutron flux simulations will be completed. The experimental results of the first qualification campaign are presented.}


\proceeding{TWEPP 2016 - Topical Workshop on Electronics for Particle Physics\\
  26-30 September 2016\\
  Karlsruhe Institute of Technology (KIT)}

\begin{document}
\maketitle
\flushbottom

\section{Introduction}
\label{sec:intro}

The Mu2e Experiment \citep{MU2ETDR} will search for coherent, neutrinoless conversion of muons into electrons in the field of a nucleus. Even if no signal is observed, the current experimental limit on the ratio between the conversion rate and the capture rate will be improved up to R$_{μe}$ <6 x 10$^{-17}$ (90$\%$ C.L.). Mu2e will then be able to probe energy scales up to thousands of TeV, far above the highest energy reachable at the most powerful colliders. \\

\begin{figure}[htbp]
\centering 
\includegraphics[width=1.\textwidth]{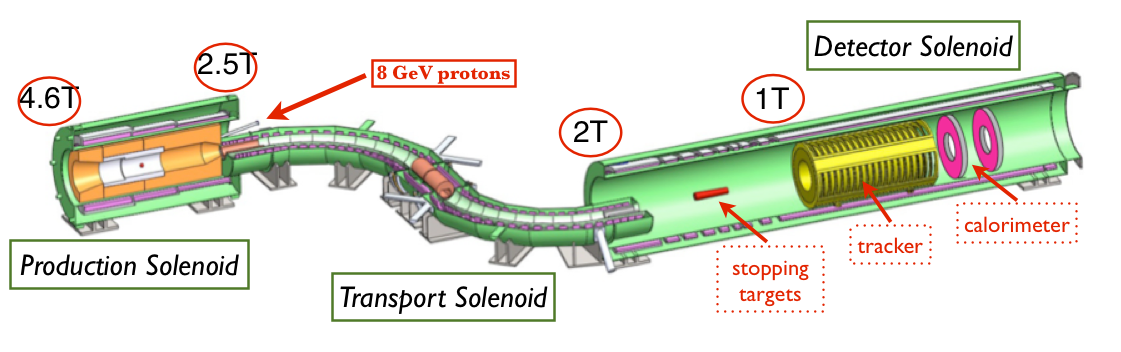}
\caption{\label{fig:beam} Diagram of the Mu2e muon beam-line and detector.}
\end{figure}

\noindent The apparatus consists of three superconductive solenoids arranged as in Figure \ref{fig:beam}: a pulsed 8~GeV proton beam hits the Tungsten Production Target inside the Production Solenoid (PS); thanks to the graded magnetic field, a large fraction of the $\pi$ and K mesons produced are directed to the Transport Solenoid, where the negative particles of low momentum are selected; the muons coming from the $\pi$ and K mesons decays are eventually stopped in the Aluminum Stopping Target in the Detector Solenoid (DS) where they form muonic atoms with a mean lifetime of 864~ns. The signal converted  electron (CE) from the conversion process $\mu^{-}+Al\to e^{-}+Al$ has a monochromatic energy of 104.96~MeV, while all the other muon decays produce background particles with lower energy. The graded field in the DS directs the electrons produced by the stopped muons decays or conversions towards the detectors: a straw tube tracker and an electromagnetic calorimeter. Both detectors are in a 10$^{-4}$~Torr vacuum and in a 1~T  solenoidal field. The tracker has to reconstruct the CE momentum with an impressive resolution of $\sim$200~keV/c. The calorimeter complements the tracker providing a powerful particle identification, a seed for the pattern recognition in the tracker and an independent software trigger. The requirements for the calorimeter are: a time resolution of better than 500~ps and an energy resolution of O(7$\%$) at the CE energy of $\sim$100~MeV. 

\section{The Calorimeter Waveform Digitizer}
\label{sec:digi}

\begin{figure}[htbp]
\centering 
\includegraphics[width=0.95\textwidth]{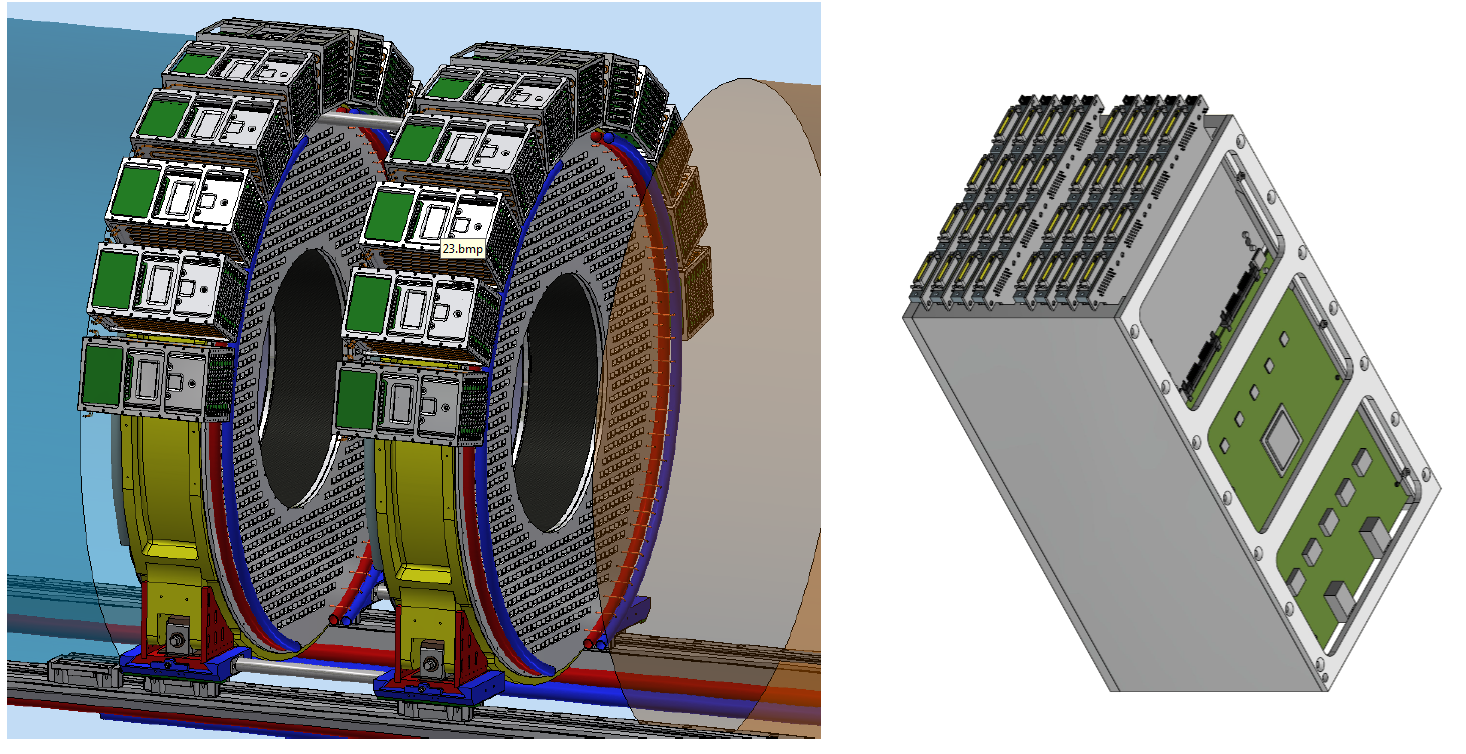}
\caption{\label{fig:calo} Left: CAD drawing of the Mu2e calorimete, consisting of an  array of CsI crystals arranged in two annular disks; electronic crates are arranged on the top of the external radii. Right: CAD drawing of a digitizer crate.}
\end{figure}

The calorimeter \cite{statuscal} is a high-granularity crystal calorimeter consisting of 1348 undoped CsI crystals, arranged in two disks as in Figure \ref{fig:calo}. Each crystal is coupled to two large-area UV-extended SiPMs arrays. Each array, also defined as calorimeter channel, consists of two series of 3 SiPMs. The two series are connected in parallel by the front-end electronics to have a x2 redundancy.\\

\noindent Since a very intense particle flux is expected to hit the calorimeter, a high sampling rate is needed to digitize the fast SiPMs signal with enough samples to precisely reconstruct the rising edge and to achieve a good resolution of the pileup of incoming particles. A waveform digitizer board, serving 20 calorimeter channels with 200 MHz 12-bit ADCs, is currently being designed. The whole system will include 140 boards to read out the 2696 calorimeter channels.\\    

\noindent The digitizers crates will be hosted inside the magnet cryostat to limit the number of pass-through connectors, disposed around the external radius of the calorimeter disks as shown in Figure \ref{fig:calo}. This complicates the design in terms of available space, power dissipation and accessibility in case of failures. Moreover, the digitizer boards have to stand all the radiation coming directly from the interactions in the stopping target, mainly bremsstrahlung photons from the beam flash and neutrons from muon captures. Most of the charged particles are trapped at a lower radius by the solenoidal field and cannot reach the electronics crates. 
Initial radiation simulations made with GEANT4/MARS estimates that the digitizer components will absorb a total ionizing dose (TID) of approximately 1.5~krad/year \cite{vitaly} and will suffer a displacement damage equivalent to the one produced by a flux of 0.5$\times$10$^{11}$ 1~MeV (Si) neutrons/cm$^{2}$ \cite{vitaly}. The simultaneous presence of radiation and high static magnetic field imposes stringent constraints which can, nonetheless, be satisfied by commercial devices, even including a safety factor.\\

\begin{figure}[htbp]
\centering 
\includegraphics[width=.65\textwidth]{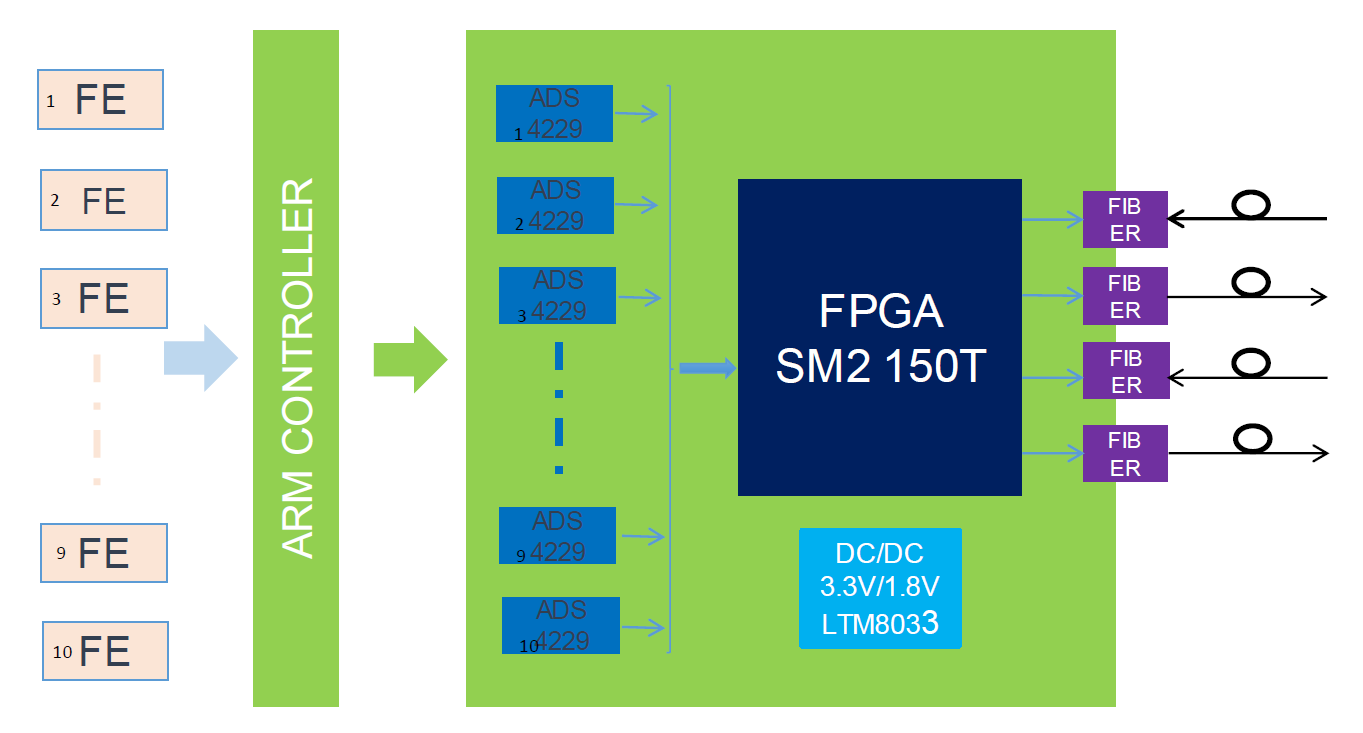}
\caption{\label{fig:digischeme} Block diagram for the calorimeter waveform digitizer.}
\end{figure}

\smallskip \noindent Each board will receive 20 amplified and shaped signals from SiPM arrays. Signals will be sampled by 10 double-channels high-speed ADCs. Their differential outputs will be connected in parallel to an FPGA. The FPGA will perform zero suppression, package the ADC information and send it to the DAQ servers through an optical link. All the DC low voltages needed by the digitizer board are locally produced starting from an external 28~V voltage source by means of DC/DC converter modules. The choice of the DC/DC converter and the ADC is critical for the project and needed a dedicated measurement campaign to test radiation hardness and operation in high magnetic field. Since the firmware is not yet defined, the qualification of the FPGA will be object of a subsequent test campaign.  

\section{The Qualification Campaign}
\label{sec:camp}

Understanding of radiation effects on silicon devices has an impact on their design and allows the prediction of a specific device behavior when exposed to a radiation field of interest. Damage inflicted to the electronic devices by a single particle (single event effect) can be temporary or permanent. It is treated separately from the TID effect for which the accumulated fluence causes degradation and from the displacement damage induced by the non-ionizing energy-loss (NIEL) deposition. Temporary effects like bit flipping (Single Event Upset) are more related to digital or mixed logic. Many tests are required to verify whether a device can be used in the Mu2e scenario; in particular the stability of the DC/DC converter and the ADC \footnote{Also the CsI crystals and the SiPMs have been tested: results can be found in \cite{Sipmandcry}.} with respect to ionizing dose, neutron flux and magnetic field have been checked in online mode, operating the devices while irradiating. To determine the integrated TID and neutron flux needed in the qualification test we started from the still rough simulated values in GEANT4/MARS \cite{vitaly} and we multiplied them by a safety factor. The final chosen numbers were a TID of 19.2 krad and a neutron flux of 8.4 $\times$ 10$^{11}$ 1~MeV (Si) n/cm$^{2}$. \\

\noindent Ionizing dose irradiation was performed at the CALLIOPE facility \cite{calliope} of the ENEA in Bracciano, where a $^{60}$Co source with an activity of 0.35$\times$10$^{15}$~Bq allowed to reach a TID 0.5~krad/h with the devices at 5~m distance. The neutron irradiation was performed at the FNG facility of the ENEA in Frascati \cite{fng}, locating the devices 6~cm far from the 14~MeV neutron source and obtaining a local flux of approximately 10$^{8}$ 1~MeV (Si) n/cm$^{2}$/s. Magnetic exposure has been performed at the LASA facility of INFN in Milan, that can supply a variable magnetic field up to 1.2~T. \\ 

\subsection{DC/DC converter tests}
\label{sec:dcdc}

Because of the presence of a strong magnetic field, the most critical component of the project is the DC/DC converter. A large variety of commercial devices from Linear Technologies have been evaluated and finally the LTM8033 has been selected, due to \cite{dcdc_mag, dcdc_doseneu}. The LTM8033 is an EMI-compatible 36~V, 3~A  DC/DC $\mu$Module buck converter. It has the interesting feature to have all the magnetics integrated inside the metal package and with an additional magnetic shielding to suppress RF interference. The device has been tested using the evaluation board DC1623A. It is pre-configured for a 3.3~V output from a 5.5~V to 36~V input. The module was powered by 36~V during the tests, with a resistive load of 1~$\Omega$ on the output. The stability of the device has been continuously monitored for the entire duration of the tests. The setup was composed of an automated system capable of measuring and storing the input/output voltages and currents every 0.5~s.\smallskip

\begin{figure}[htbp!]
\centering 
\includegraphics[width=0.8\textwidth]{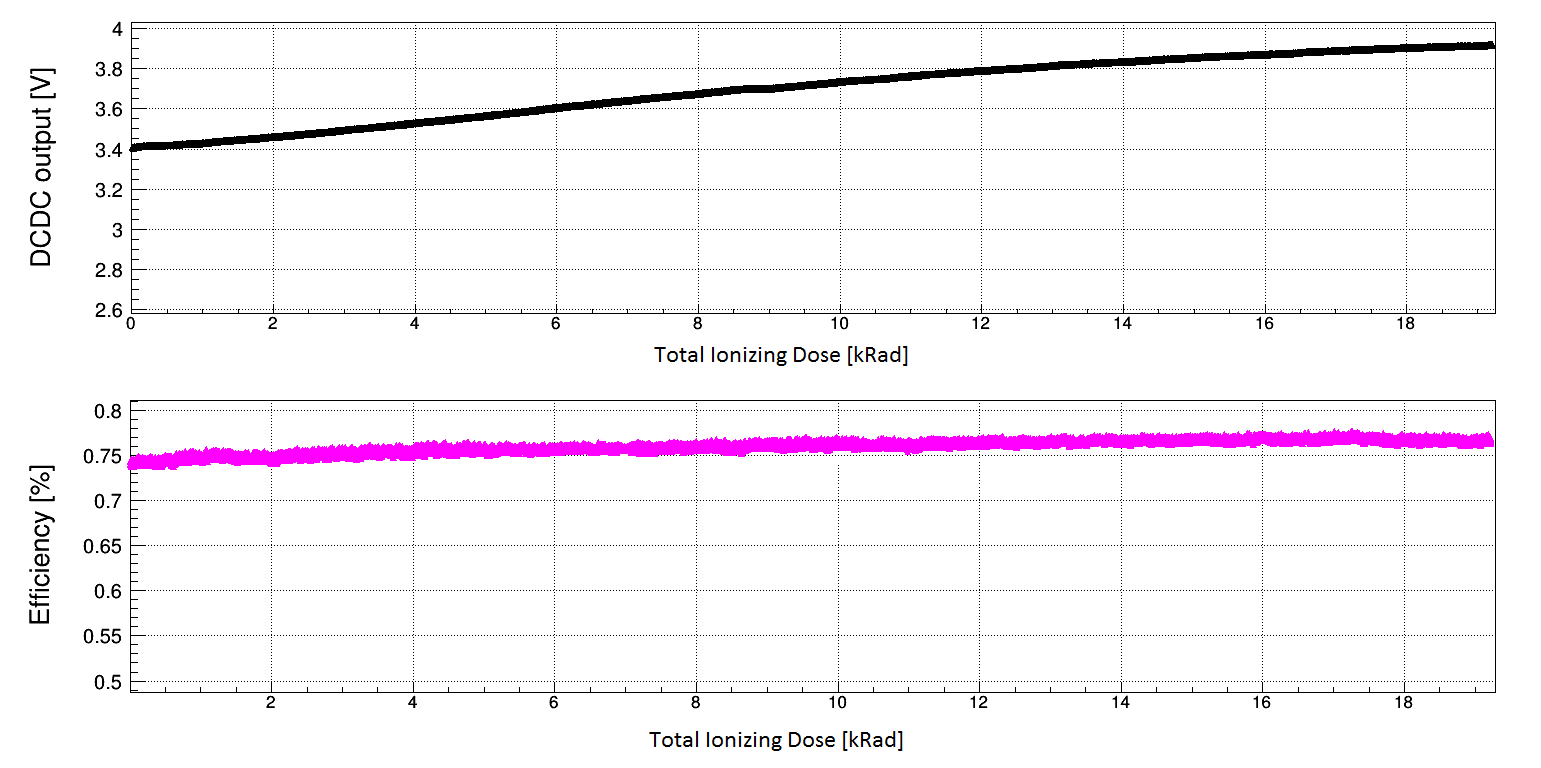}
\caption{\label{fig:dcdose} The output voltage (up) and efficiency (down) of the LTM8033 as a function of the TID up to 20~krad.}
\end{figure}

\paragraph{Ionizing Dose} - The module has been tested up to 20~krad, as shown in Figure \ref{fig:dcdose}. We observed an increase of the output voltage of about 0.5~V at 20 krad, $\sim$70~mV at the TID expected for Mu2e. \smallskip

\begin{figure}[htbp!]
\centering 
\includegraphics[width=0.8\textwidth]{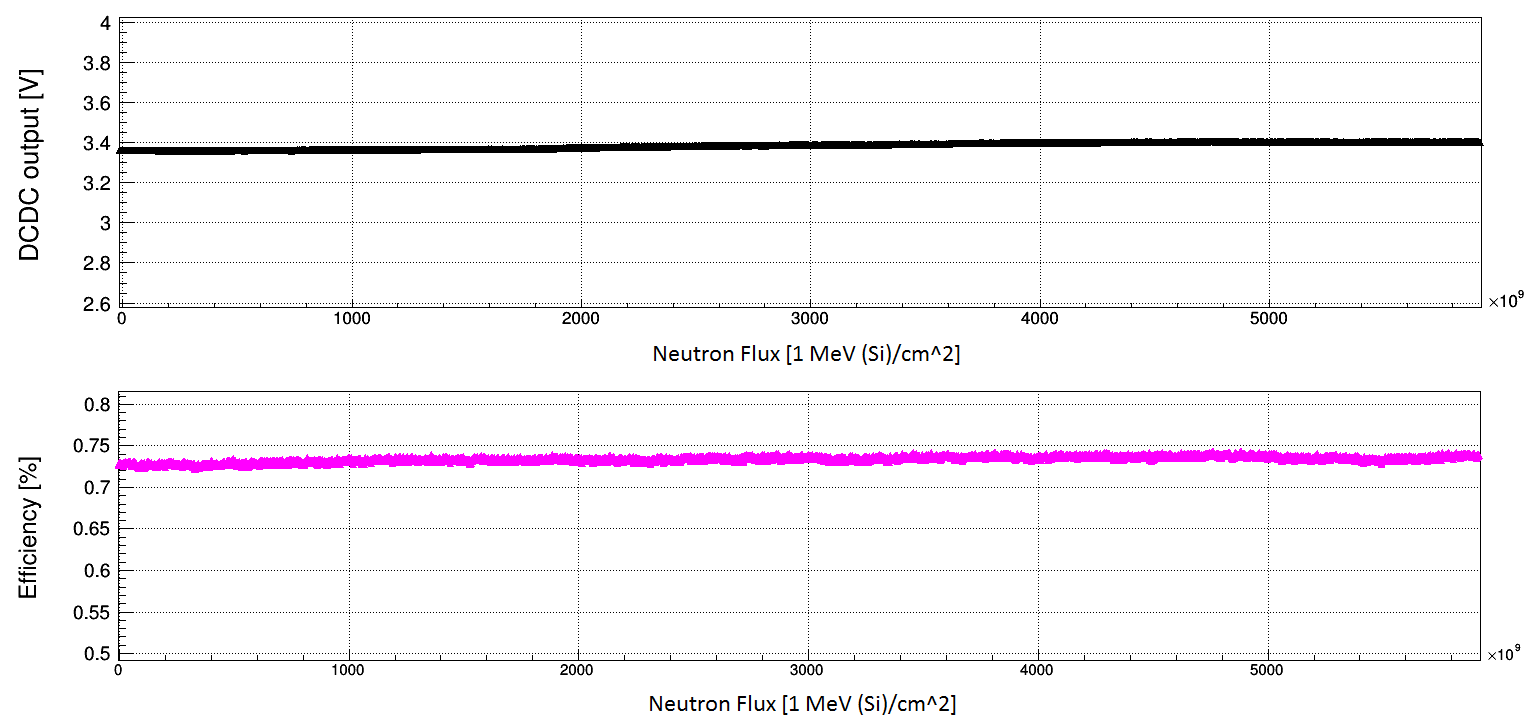}
\caption{\label{fig:dcneu} The output voltage (up) and efficiency (down) of the LTM8033 as a function of the equivalent displace damage in 1 MeV (Si) n/cm$^{2}$.}
\end{figure}

\paragraph{Displacement Damage} - After an irradiation of 5$\times$10$^{12}$ 1 MeV (Si) n/cm$^{2}$ only a 40~mV increase in the output voltage has been observed, see Figure \ref{fig:dcneu}. 

\begin{figure}[htbp]
\centering 
\includegraphics[width=.68\textwidth]{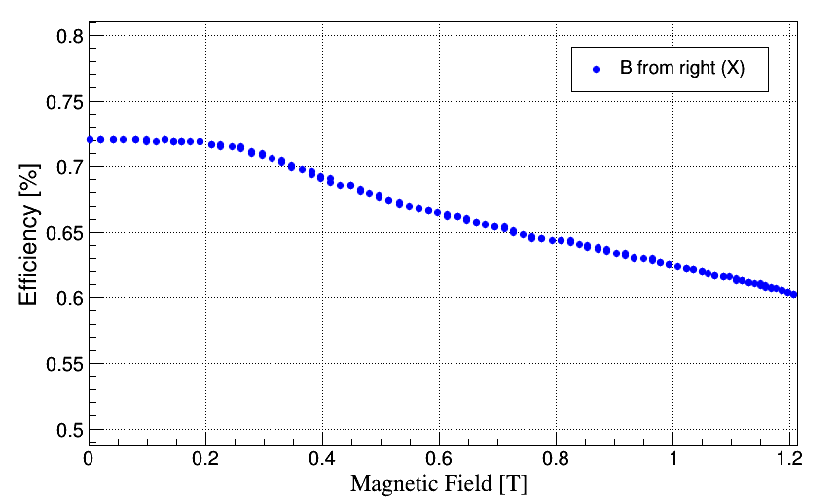}
\quad
\includegraphics[width=.27\textwidth]{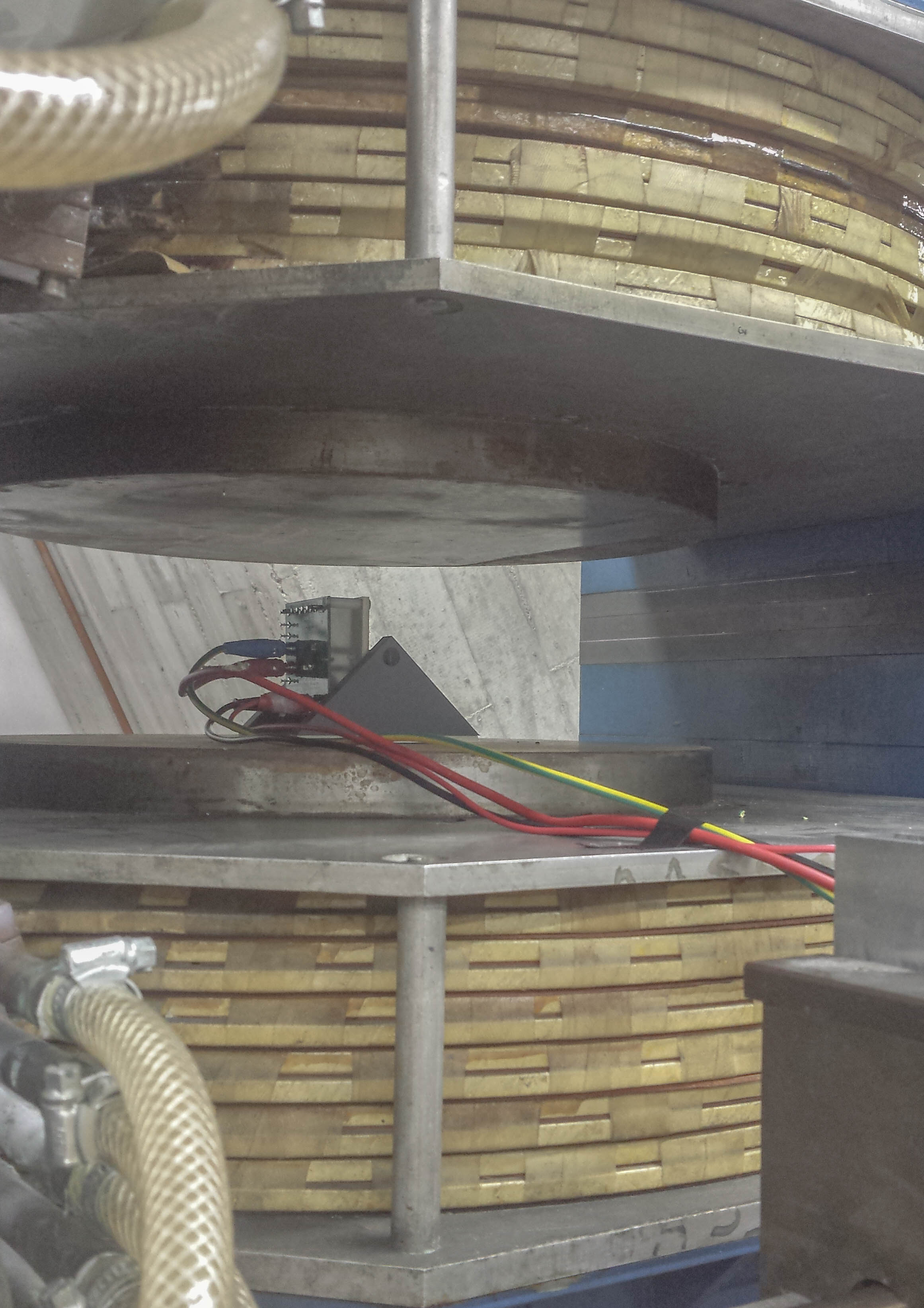}
\caption{\label{fig:lasa} Left: The efficiency of the LTM8033 packaged module as a function of the magnetic field intensity. Right: The evaluation board under test inside the magnet at LASA.}
\end{figure}

\paragraph{Magnetic Field} - The module has been exposed to B-field in three different orientations without significant changes. Figure \ref{fig:lasa} shows the case with the field on the x-axis \footnote{The magnetic field is perpendicular to the short side of the chip.} of the evaluation module: the input current tends to increase slightly with the increase of the applied magnetic field, with a consequent decrease in conversion efficiency starting from 0.2~T. The device is still providing the 3.3 V output voltage at 1~Tesla, despite a total drop in efficiency of $\sim$10$\%$.

\subsection{ADC tests}
\label{sec:adc}

The choice of the ADC has been guided by the following requests: a sampling frequency of at least 200 MHz, a resolution of 12-bit, low power consumption to operate in vacuum and low cost. The Texas Instrument ADS4229 has been selected as the best value for money. It is a commercial dual-channel, 12 bit, 250-MSPS ultralow-power ADC with parallel LVDS output. The ADC performance has been monitored using a custom setup. A 200~kHz standard sinusoidal, tuned to fulfill the ADC dynamic scale, has been used as input for both the ADC channels. Instead of store digital data we used a custom DAC board, located far from the radiation zone, to compare the ADC output with the analog sinusoidal wave. The DAC output waveforms has been acquired through a digital oscilloscope (50~$\mu$s samples with a 10 Hz frequency), up to 20~krad and 6$\times$10$^{12}$ 1~MeV~(Si)~n/cm$^{2}$. Gathering more than 300~GB of data from both the tests, no significant changes in the analog output or bit flips have been observed.  

%

\section{Conclusions}
\label{sec:digi}

Two of the critical components of the Mu2e calorimeter waveform digitizer board have been selected and individually qualified towards currently known limits: DC/DC converter and ADC have been qualified to operate in high magnetic field and to survive to a ionization dose of 20 krad and a neutron flux of 6 $\times$~10$^{12}$~1~MeV~(Si)~n/cm$^2$. The design of the first prototype is almost completed and is expected to be ready for construction in early 2017. All the qualification tests need to be repeated on the prototype before freezing the design, when the environment simulations will be validated.

\acknowledgments

This work was supported by the EU Horizon 2020 Research and Innovation Programme under
the Marie Sklodowska-Curie Grant Agreement No.  690835.

\end{document}